\documentstyle [12pt] {article}

\catcode`@=12
\begin{document}
\thispagestyle{empty}
\begin{flushright} UCRHEP-T236\\
DTP/98/62\\
October 1998\
\end{flushright}
%\mbox{}
\vspace{0.2in}
\begin{center}
{\Large	\bf A See-Saw Model for Atmospheric and Solar\\
Neutrino Oscillations \\}
\vspace{0.5in}
{\bf Ernest Ma$^1$, D.P. Roy$^{2,3,4}$, and Utpal Sarkar$^{2,5}$\\}
\vspace{0.2in}
{$^1$ \sl Department of Physics, University of California\\}
{\sl Riverside, California 92521, USA\\}
\vspace{0.1in}
{$^2$ \sl DESY, Notkestrasse 85, D-22607 Hamburg, Germany\\}
\vspace{0.1in}
{$^3$ \sl Department of Physics, University of Durham\\}
{\sl Durham DH1 3LE, UK\\}
\vspace{0.1in}
{$^4$ \sl Tata Institute of Fundamental Research,\\} 
{\sl Homi Bhabha Road, Mumbai 400 005, India\\}
\vspace{0.1in}
{$^5$ \sl Physical Research Laboratory, Ahmedabad 380 009, India\\}
\vspace{0.5in}
\end{center}
\begin{abstract}\

We have constructed an explicit see-saw model containing two singlet 
neutrinos, one carrying a $(B-3L_e)$ gauge charge with an intermediate
mass scale of $\sim O(10^{10})$ GeV along with a sterile one near the
GUT (grand unification theory) scale of $\sim O(10^{16})$ GeV. With these 
mass scales and a reasonable range of Yukawa
couplings, the model can naturally account for the near-maximal mixing
of atmospheric neutrino oscillations and the small mixing
matter-enhanced oscillation solution to the solar neutrino deficit. 

\end{abstract} 

\newpage
\baselineskip 18pt

The super-Kamiokande experiment has recently provided convincing 
evidence for the atmospheric neutrino oscillation \cite{atm} as well 
as confirmed earlier results on
solar neutrino oscillation \cite{sol}. The atmospheric neutrino 
oscillation data seem to require a large mixing angle between
$\nu_\mu$ and $\nu_\tau$,
\begin{equation}
\sin^2 2 \theta_{\mu \tau} > 0.82
\end{equation}
and 
\begin{equation}
\Delta M^2 = (0.5-6) \times 10^{-3} {\rm eV}^2.
\end{equation}
On the other hand, the solar neutrino oscillation data can be explained 
by the small mixing-angle matter-enhanced solution between $\nu_e$ and a 
combination of $\nu_\mu/\nu_\tau$ with \cite{sam}
\begin{equation}
\sin^2 2 \theta_{e-\mu/\tau} = 10^{-2} - 10^{-3}
\end{equation}
and 
\begin{equation}
\Delta m^2 = (0.5-1) \times 10^{-5} {\rm eV}^2.
\end{equation}
This represents the most conservative solution to the solar neutrino 
anomaly although one can get equally good solutions with large 
mixing-angle matter-enhanced and vacuum oscillations as well. 
One would naturally 
expect a near-maximal mixing between $\nu_\mu$ and $\nu_\tau$ (1),
as required by the atmospheric neutrino data, if they were almost
degenerate Dirac partners with a small mass difference given by
(2). In the context of a three-neutrino model however, the solar
neutrino solution (4) would then require the $\nu_e$ to show a 
much higher level of degeneracy with one of these states, which
is totally unexpected. Therefore, it is more natural to consider the
three neutrino mass states as nondegenerate with
\begin{equation}
m_1 = (\Delta M^2)^{1/2} \simeq 0.05 {\rm eV}, ~~~~ m_2 = (\Delta m^2)^{1/2}
\simeq 0.003 {\rm eV}, ~~~~m_3 << m_2.
\end{equation}
There is broad agreement on this point in the current literature on
neutrino physics \cite{4}, much of which is focussed on the question of 
reconciling this hierarchical structure of neutrino masses with at
least one large mixing angle (1). 

The cannonical mechanism for generating neutrino masses and mixings is 
the so called see-saw model involving heavy right-handed singlet
neutrinos \cite{seesaw}. It naturally leads to small hierarchical 
masses for the three doublet neutrinos, but with small mixing angles. 
Alternatively one can generate the small neutrino masses radiatively
via the Zee model \cite{zee,ma} or the R-parity breaking supersymmetric 
model \cite{susy}. Instead of heavy right-handed neutrinos, one needs
here an expanded scalar sector in the $\leq$ TeV region, as extra
Higgs multiplets in the former case and as squarks and sleptons in the 
latter. The radiative mechanism offers more flexibility to reconcile
hierarchical neutrino masses with at least one large mixing angle. In fact,
explicit models for neutrino masses and mixing have been constructed 
recently to explain the atmospheric and solar neutrino data in terms of these 
two radiative mechanisms \cite{ma,susy}. It should be noted however that
the presence of extra scalars in the $\leq$ TeV region in these models
represents a potential problem with large flavour-changing neutral-current 
(FCNC) effects. Moreover, these extra scalars can either be detected 
or ruled out in future colliders. On the other hand, the see-saw model 
is less vulnerable to FCNC effects and collider search, although it 
is harder to reconcile hierarchical neutrino masses with a large 
mixing in this case. The present work is devoted to this 
exercise. As we shall see below, this model can naturally reconcile
hierarchical neutrino masses with a large mixing angle (1). Moreover,
the low-energy ($\leq$ TeV) spectrum of this model is identical
to the standard model, so that it has no potential problem with
flavour-changing neutral currents.

Let us first consider the atmospheric neutrino oscillation. It is 
clear from (1), (2) and (5) that it requires the heaviest neutrino 
state to be a roughly equal mixture of $\nu_\mu - \nu_\tau$ with 
mass $\sim 0.05$ eV. In the simplest see-saw model, this requires one 
heavy singlet neutrino, having a Dirac coupling to this equal mixture of
$\nu_\mu - \nu_\tau$ \cite{king}. Such a heavy neutrino can be motivated
in a U(1) extension of the standard model, with the U(1) gauge charge
corresponding to $(B-3 L_i)$, where one needs one right-handed singlet 
neutrino carrying an $L_i$ number of 1 for anomaly cancellation \cite{ma1}.
This is analogous to the left-right symmetric model, corresponding
to the U(1) gauge charge $(B-L_e-L_\mu-L_\tau)$, where one needs three 
right-handed neutrino singlets for anomaly cancellation. Such a 
U(1) extension of the standard model was recently constructed for the 
U(1) gauge charge $(B-3L_\tau)$ \cite{ma1} and its phenomenological
implications studied \cite{dp}. Although one can make the right-handed 
singlet neutrino of this model to couple to a roughly equal mixture of 
$\nu_\mu - \nu_\tau$ by adjusting the model parameters, it will not
be a natural feature of this model. To achieve this naturally, we
must treat the $\mu$ and $\tau$ flavours on equal footing and 
distinguish them from $e$. Accordingly we shall consider the U(1) extension 
of the standard model, corresonding to the U(1) gauge charge,
\begin{equation}
Y' = B - 3 L_e .
\end{equation}
Moreover, we shall introduce a reflection symmetry via a multiplicative 
quantum number, N--parity, in order to avoid the coupling of the singlet
neutrino with $\nu_e$ \cite{babu}. 

The leptons and Higgs scalars of the model are listed below with their 
$SU(2)_L \times U(1)_Y \times U(1)_{Y'}$ gauge charges, where the 
negative N--parity states have been identified by the subscript. 
\begin{eqnarray}
\pmatrix{\nu_e \cr e}_L \sim (2,-1/2;-3) &\hskip .5in&
{\nu_e^c}_L \sim (1,0;3)_{\_} \nonumber \\
\pmatrix{\phi^+ \cr \phi^0}_L \sim (2,1/2;0) &\hskip .5in&
\pmatrix{\eta^+_1 \cr \eta_1^0}_L \sim (2,1/2;3) \nonumber \\
\pmatrix{\eta^+_2 \cr \eta_2^0}_L \sim (2,1/2;-3)_{\_} &\hskip .5in&
\chi^0 \sim (1,0;-6) \nonumber \\
\zeta^0 \sim (1,0;-3) &\hskip .5in&
\nu_S \sim (1,0;0) 
\end{eqnarray}
Here one extra singlet neutrino $\nu_S$ with no $Y'$ charge has been added
for explaining the solar neutrino data. Since the mass $M$ of this 
sterile neutrino is not protected 
by any symmetry, it is expected to be very high, going up to the 
GUT scale ($10^{16}$ GeV). All the other new particles will acquire
masses at an intermediate scale, corresponding to the spontaneous 
breaking of the $U(1)_{Y'}$ gauge symmetry.
The resulting mass matrix for the five neutrino states in the
basis $[\nu_e ~~ \nu_\mu ~~ \nu_\tau ~~ \nu_e^c ~~ \nu_S]$ is given by,
\begin{eqnarray}
\pmatrix{0&0&0&0& f'' <\eta_1> \cr
 0&0&0& f_1 <\eta_2>& f_1' <\phi> \cr
 0&0&0& f_2 <\eta_2>& f_2' <\phi> \cr
 0&  f_1 <\eta_2>& f_2 <\eta_2>& f <\chi> &0\cr
 f'' <\eta_1>&f_1' <\phi> & f_2' <\phi> &0&M}
\end{eqnarray}
Note that the scalar $\zeta^0$ does not contribute to the mass matrix.
However, it provides a soft N--parity breaking term in the Lagrangian,
which allows the model to avoid a potential domain-wall problem. 

Both $\zeta^0$ and $\chi^0$ are expected to acquire large vacuum
expectation values and masses at the scale of $U(1)_{Y'}$ breaking.
In contrast, the $SU(2)$ doublets $\eta_1$ and $\eta_2$ are required
to have positive mass-squared terms at this scale, so that they would 
have large masses but very small $vev$s $<\eta_1>$ and $<\eta_2>$
\cite{prl}. For example, $<\eta_2>$ can be estimated from the 
relevant part of the scalar potential,
\begin{equation}
m_2^2 \eta_2^\dagger \eta_2 + \lambda (\eta_2^\dagger \eta_2)
(\zeta^\dagger \zeta) + \lambda' (\eta_2^\dagger \eta_2)
(\chi^\dagger \chi) - \mu \phi^\dagger \eta_2 \zeta^\dagger
\end{equation}
where the last term is the N--parity breaking soft term mentioned
above. Although we start with a positive mass-squared term
for the field $\eta_2$, after minimisation of the potential we find
that this field acquires a small non-zero $vev$ given by,
\begin{equation}
<\eta_2> = {\mu <\phi> <\zeta> \over M_2^2} 
\end{equation}
where $M_2^2 = m_2^2 + \lambda <\zeta^0>^2 + \lambda' <\chi^0>^2$
represents the physical mass of $\eta_2$ and $<\phi> \simeq 10^2$ GeV. 
One expects a similar value for $<\eta_1>$.
The size of the soft term can be anywhere up to the spontaneous
symmetry breaking scale, {\it i.e.}, $\mu \leq M_2$.

In order to account for the desired neutrino masses and mixing we 
shall require the size of the $vev$s to be 
\begin{equation}
<\eta_1> \sim <\eta_2> \sim 1 {\rm GeV}  . 
\end{equation}
This would correspond to assuming $\mu \sim <\zeta>/100$ in (10).
Alternatively one can get this with $\mu \sim <\zeta>$ and 
$M_2 \simeq m_2 \simeq 10 <\zeta>$. In either case one can get
the required $vev$ with reasonable choice of the mass parameters
around the scale of the spontaneous symmetry breaking.

We shall now proceed to calculate the masses and mixing angles of
the three light left-handed neutrinos by diagonalising the $5 \times 5$
mass matrix. Since we have added two singlet neutrinos, one of the 
doublet neutrinos will remain massless. This is also clear from the fact 
that the determinant of the mass-matrix (8) is zero.  Let 
\begin{eqnarray}
a_1 = { f_1 <\eta_2> \over \sqrt{f <\chi>}}, &\hskip 1in&
a_2 = {f_2 <\eta_2> \over  \sqrt{f <\chi>}}, \nonumber \\
b_1 = {f_1' <\phi>  \over  \sqrt{M}}, &\hskip 1in&
b_2 = {f_2' <\phi> \over  \sqrt{M}} \nonumber \\ 
&\displaystyle{c = {f'' <\eta_1> \over \sqrt{M} } }& .
\end{eqnarray}
We then take the approximation $a_{1,2} \gg b_{1,2} \gg c$, which will be 
true for our parameter space of interest. The two nonzero light mass 
eigenvalues are now 
\begin{eqnarray}
m_1 &\simeq& a_1^2 + a_2^2 , \\
m_2 &\simeq& {(a_1 b_2 - a_2 b_1)^2 \over a_1^2 + a_2^2} .
\end{eqnarray}
The $f'_{1,2}$ are Yukawa couplings of the standard model Higgs 
boson to $\nu_\mu, \nu_\tau$. Assuming them to be similar in size to
the top quark Yukawa coupling as in SO(10) grand unified theories implies
\begin{equation}
f'_{1,2} \sim 1 .
\end{equation}
On the other hand, assuming them to be similar in size to the $\tau$
Yukawa coupling would imply 
\begin{equation}
f'_{1,2} \sim 10^{-2} .
\end{equation}
Comparing (5), (12), and (14), we see that the Yukawa couplings of (15) give 
\begin{equation}
M \sim 10^{16} {\rm GeV} (\sim M_{GUT}), 
\end{equation}
while the Yukawa couplings of (16) imply
\begin{equation}
M \sim 10^{12} {\rm GeV} ,
\end{equation}
which is also a reasonable value. 

Thus one can get the right mass for the matter-enhanced solution to solar 
neutrino oscillations for $M$ in the range of $10^{12-16}$ GeV. Moreover
we see from (12) and (13) that 
\begin{equation}
m_1 \sim  \left( {f_1^2 + f_2^2 \over f <\chi>} \right) {\rm GeV} .
\end{equation}
Here the Yukawa couplings appearing in the numerator and denominator 
correspond to the scalars $\eta_2$ and $\chi^0$ respectively.
Assuming them to be of similar size, we see that any value of this 
Yukawa coupling in the range of (15) -- (16) will give the required 
$m_1$ of equation (5) for
\begin{equation}
<\chi> \sim <\zeta> \sim 10^{8-10} {\rm GeV}.
\end{equation}
Thus we can have the right mass for 
atmospheric neutrino oscillations for a reasonable scale of the 
$U(1)_{Y'}$ symmetry breaking and a reasonable range of the 
Yukawa couplings.

Let us now look at the mixing matrix connecting the neutrino flavour
eigenstates $(\nu_e, \nu_\mu, \nu_\tau)$ to the mass eigenstates
$(\nu_3, \nu_2, \nu_1)$, written in increasing
order of mass.  Because of the structure of (8) with a guaranteed zero mass 
eigenvalue, we can express this mixing matrix as a
product of two matrices $U_1$ and $U_2$ corresponding to the atmospheric
and solar neutrino mixing angles respectively, {\it i.e.}, 
\begin{equation}
\pmatrix{\nu_e \cr  \nu_\mu \cr  \nu_\tau} = U_1 U_2 
\pmatrix{\nu_3 \cr  \nu_2 \cr  \nu_1} .
\end{equation}
We get
\begin{equation}
U_1 = \pmatrix{ 1 & 0 & 0 \cr 0 & \cos \theta_1 & \sin \theta_1 \cr
0 & - \sin \theta_1 & \cos \theta_1 },
\end{equation}
where 
$$ \tan \theta_1 = {\sin \theta_1 \over \cos \theta_1 }
\simeq {a_1 \over a_2} 
\left( = f_1 \over f_2 \right) .  $$
Note that $\tan \theta_1$ is simply the ratio of the Yukawa coulings of 
$\eta_2$ to $\nu_\mu$ and $\nu_\tau$ which are expected to be of
similar size. Assuming them to be equal implies $\tan \theta_1 = 1$;
{\it i.e.}, maximal mixing for atmospheric neutrino oscillation, 
$\sin ^2 2 \theta_1 = 1$. Moreover any value of this ratio in the
range
\begin{equation}
0.64 < {f_1 \over f_2} < 1.56
\end{equation}
will ensure the near-maximal mixing condition of equation (1). Thus 
we can get the required mixing angle for atmospheric neutrino
oscillations without any fine tuning of the Yukawa couplings.

Finally we get
\begin{equation}
U_2 = \pmatrix{ \cos \theta_2 & -\sin \theta_2 
& 0 \cr \sin \theta_2 & \cos \theta_2 & 0 \cr
0 & 0 & 1 },
\end{equation}
where $$\sin  \theta_2 = {c \sqrt{a_1^2 + a_2^2} \over a_1 b_2 - a_2 b_1}
\simeq {c \over b_{1,2}} .$$ 
Substituting (12) in (24) gives
\begin{equation}
\sin \theta_2 = {f'' \over  f'_{1,2}} {<\eta_1>  \over  <\phi>} .
\end{equation}
This is to be compared with the required angle (3), which 
corresponds to 
\begin{equation}
\sin \theta_2 = ( 1.6 - 5) \times 10^{-2} .
\end{equation}
Assuming the Yukawa couplings to be of similar size, equation
(25) gives the required mixing angle for 
\begin{equation}
<\eta_1> \sim 1 {\rm GeV}
\end{equation}
as mentioned earlier. There is a contribution to this angle from
the charged lepton sector, which is however relatively small, as we
see below. 

We have been working in the basis where the charged-lepton mass matrix,
arising from their couplings to the standard-model Higgs boson $\phi$, is 
diagonal. However, there will be non-diagonal terms introduced by the 
Yukawa couplings of $\eta_1$ to $e \mu$ and $e \tau$ \cite{ma};
{\it i.e.}
\begin{equation}
M_\ell = \pmatrix{ m_e & 0 & 0 \cr
f''_1 <\eta_1> & m_\mu & 0 \cr
f''_2 <\eta_1> & 0 & m_\tau }
\end{equation}
Its contribution to the $\nu_e$ mixing angle is
\begin{equation}
\sin \theta_2 = f''_1 <\eta_1> {m_e \over m_\mu^2} \sim
f''_1 ~ 10^{-1} \leq 10^{-3},
\end{equation}
since the Yukawa coupling in this case is at most $\sim 10^{-2}$.

In summary, we have constructed a see-saw model containing two heavy 
singlet neutrinos. One of them carries a $(B-3L_e)$ gauge charge and 
acquires an intermediate scale mass, corresponding
to the spontaneous breaking of this gauge symmetry. The associated 
scalars have also masses at this scale. The other singlet neutrino is 
sterile and has a very heavy mass near the GUT scale. With these two mass
scales and a reasonable range of 
Yukawa couplings, the model can naturally
account for the near maximal mixing solution to the atmospheric
neutrino oscillation and the small mixing MSW solution to the solar
neutrino oscillation. 

~\vskip 0.5in
\begin{center} {ACKNOWLEDGEMENT}
\end{center}

We thank M. Drees for discussions.
We acknowledge the 
hospitality of the DESY Theory Group and US acknowledges financial 
support from the Alexander von Humboldt Foundation. The work of EM was 
supported in part by the U.~S.~Department of Energy under Grant 
No.~DE-FG03-94ER40837.

\newpage
\bibliographystyle{unsrt}

\end{document}